\documentclass[12pt]{article}
\usepackage{amsmath,XoohmE}
\usepackage{longtable}
\SfTitles 
\seceq
\allowdisplaybreaks
\def\bs{\sf\bfseries\boldmath}
\def\Aut{\mathop{\text{Aut}}\nolimits}
\def\Tr{\mathop{\text{\large Tr}}}
\def\TM{$^{\sss\text{TM}}$}
\def\cc{(0)}

\def\op{\oplus}
\def\TO{\twoheadrightarrow}
\begin{document}
\thispagestyle{empty}
\begingroup
 \renewcommand{\thefootnote}{\alph{footnote}}
\vglue0mm\vfill
\begin{center}
{\LARGE\bs
 $\ZZ_7$ Orbifold Models in M-Theory}\\[5mm]
{\large\bs  M.K.~Ahsan\ft{Email: mkahsan@gmail.com} {\rm and}
            T.~H\"{u}bsch\ft{Email: thubsch@mac.com}}\\[1mm]
{\it
  Department of Physics and Astronomy,\\[-1mm]
      Howard University, Washington, DC 20059.
}\\[20mm]
{\bs ABSTRACT}\\[5mm]
\parbox{135mm}{\parskip=\medskipamount%
Among $T^7/\G$ orbifold compactifications of $M$-theory, we examine models containing the particle physics Standard Model in four-dimensional spacetimes, which appear as fixed subspaces of the ten-dimensional spacetimes at each end of the interval, $I^1\simeq S^1/\ZZ_2$, spanning the $11^\text{th}$ dimension.
 Using the $\ZZ_7$ projection to break the $E_8$ gauge symmetry in each of the four-planes and a limiting relation to corresponding heterotic string compactifications, we discuss the restrictions on the possible resulting gauge field and matter spectra.
 In particular, some of the states are non-local: they connect two four-dimensional Worlds across the $11^\text{th}$ dimension.

We illustrate our programmable calculations of the matter field spectrum, including the anomalous $U(1)$ factor which satisfies a universal Green-Schwarz relation, discuss a Dynkin diagram technique to showcase a model with $SU(3)\times SU(2)\times U(1)^5$ gauge symmetry, and discuss generalizations to higher order orbifolds.}
\end{center}


\vfill\vfill
\setcounter{footnote}{0}
\endgroup
\clearpage

\pagestyle{plain}
\setcounter{page}{1}
\section{Introduction}
\label{intro}
The strong coupling limit of type IIA string theory is dual to a theory, the low energy limit of which exhibits a target space with eleven-dimensional Lorentz invariance\cite{witten1,townsend1,polchinski1,becker1}; this still rather mysterious extension of string theory is known as $M$-theory. Unlike in string theory, not enough is known about the underlying dynamics of $M$-theory, so that much of the analysis relies on this low energy limit, which includes the eleven-dimensional $N{=}1$ supergravity. In turn, eleven is the maximum number of spacetime dimensions for a consistent locally supersymmetric theory, and has been studied independently of relations to string theory. The basic five different string theories and the eleven-dimensional supergravity are in fact all regarded as six different special locations (limits) in the moduli space of this unifying $M$-theory\cite{polchinski1,becker1}.

Just as type IIA string theory in ten-dimensional spacetime is related to $N{=}1$ supergravity in $\IR^{1,9}\times S^1$, the $E_8\times E_8$ heterotic string theory is related to supergravity in $\IR^{1,9}\times I^1$\cite{horava1}. In this latter type of models, the $11^\text{th}$ dimension spans the interval $I^1\simeq S^1/\ZZ_2$, where $\ZZ_2$ is the parity reflection.
 The boundary of this eleven-dimensional spacetime thus consists of two parallel ten-planes, located at the two orbifold singularities: the endpoints of the $11^\text{th}$ dimension. Local anomaly cancellation enforces the emergence of an $E_8$ gauge symmetry in each of the two boundary ten-planes.

Within $M$-theory compactifications of the last seven (spatial) dimensions of the $\IR^{1,10}$ spacetime on a $T^7/\G$ orbifold\ft{The torus-compactification $\IR^{1,9}\times (S^1/\ZZ_2) \to \IR^{1,3}\times T^6\times(S^1/\ZZ_2)$ is seen to result in $\IR^{1,3}\times T^7/\ZZ_2$, where $T^6\times S^1=T^7$ and the $\ZZ_2$ parity reflection acts only on the last, $11^\text{th}$ coordinate.}, where $T^7=\IR^7/\L$ and $\G\subset\Aut(\L)$, we explore the cases where $\G$ is a finite abelian group of rotations in the first six dimensions of $T^7$, twisted by the parity reflection in the last, $11^\text{th}$ dimension. 
 In general, all such actions fix entire copies of $\IR^{1,3}$ located at certain $p$-dimensional ``planes'', $P_\star\subset T^7$, fixed by the $\G$ action. Each such $P_\star$ is then a singularity in $T^7/\G$, and defines a special $(1,3{+}p)$-dimensional spacetime, $(\IR^{1,3}\times P_\star)\subset(\IR^{1,10}/\L)/\G$, called ``orbifold fixed plane'' in Refs.\cite{doran3,doran2}. There, several models were constructed using $T^6/(\ZZ_2\times \ZZ_2)\times(S^1/\ZZ_2)$ and  $T^6/(\ZZ_2\times \ZZ_3)\times(S^1/\ZZ_2)$ orbifolds, and the four-dimensional spacetime occurs where $\dim(P_*)=0$.

In principle, the physics in these orbifold fixed planes are beset with gravitational and gauge anomalies, carried by Weyl (chiral) fermions. Thus, cancellation of these anomalies plays a crucial role in studying $M$-theory phenomenology\cite{kaplunovsky1}. As there can be no Weyl fermions in odd-dimensional spacetimes, odd-dimensional orbifold fixed planes can have no such anomaly. In turn, in $4k{+}2$ dimensions, $k=0,1,2,\cdots$, there are purely gravitational anomalies due to spin-$1/2$ and spin-$3/2$ Weyl fermions. Also, in $4k{+}2$ dimensions, gauge anomaly may arise only if the left- and right-handed fermions transform differently under the gauge group.

Finally, in four-dimensional spacetime, there can be no gravitational anomaly and gauge anomaly is generated by Weyl fermions transforming in complex representations. A cancellation of these anomalies is necessary for the quantum theory to be well-defined, and leads to restrictions on the allowed gauge group representations of the theory. Generally, the four-dimensional gauge group contains one or more $U(1)$ factor, each of which may receive non-vanishing contributions to its anomaly. A basis change easily renders all but one of these $U(1)$ factors anomaly-free, but the anomalies in the last $U(1)$ factor may in general only be cancelled by a four-dimensional counterpart of the Green-Schwarz mechanism\cite{green1}.

\begin{figure}[ht]
\begin{center}
\begin{picture}(100,40)(-10,0)
\put(5,31.8){{}\llap{\fbox{$M$-theory/$\ZZ_2$}}}
\put(10,32.5){\vector(1,0){35}}
\put(20,35){$x_{11}\rightarrow 0$}
\put(50,31.8){\fbox{$E_8\times E_8$ heterotic string theory}}
\put(0,27){\vector(0,-1){20}}
\put(57,27){\vector(0,-1){20}}
\put(-2,17){{}\llap{$\ZZ_N$}}
\put(59,17){$\ZZ_N$}
\put(5,0){{}\llap{\fbox{Explored}}}
\put(10,1){\vector(1,0){35}}
\put(20,4){$x_{11}\rightarrow 0$}
\put(50,0){\fbox{Known Models}}
\end{picture}
\vspace{3mm}
\caption{Relations between $M$-theory and $E_8\times E_8$ heterotic string theory orbifold models.}
\label{ms}
\end{center}
\end{figure}
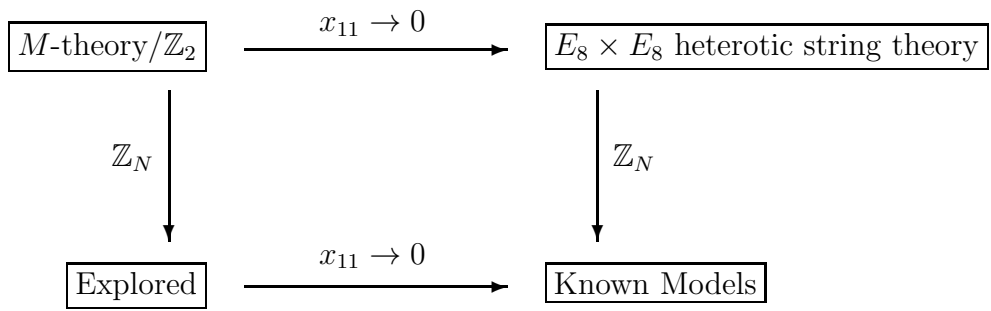

The massless spectra in such orbifold models are closely related to the corresponding models found from compactifying the $E_8\times E_8$ heterotic string theory on $T^6/\ZZ_N$ orbifolds\cite{katsuki1}; see Figure~\ref{ms}. $M$-theory compactified on $S^1/\ZZ_2$ has two fixed ten-plane boundaries with a chiral $N=1$, $E_8$ gauge vector supermultiplet in each of them. In the $x_{11}\to0$ limit, this 11-dimensional spacetime limits to a 10-dimensional spacetime---the two boundary ten-plane spacetimes coalescing---together with their separate copies of the $E_8$ gauge bundle. Without going into the details of the limiting dynamics, the massless spectrum in this limit is clearly identifiable with that in the $E_8\times E_8$ heterotic string theory. The massless spectra of the well-studied four-dimensional $E_8\times E_8$ heterotic string orbifold models then ought to agree with those in the $x_{11}\to0$ limit of corresponding orbifold compactifications of the $M$-theory.
We will see that this limiting behavior\ft{The massless spectra obtained {\em\/via\/} this limit are consistent with known anomaly cancellation results.} is usefully restrictive, although certainly not sufficient to determine the \emph{\/complete\/} spectrum. In turn, note that anomaly cancellations alone can similarly not rule out the occurrence of arbitrary multiples of collections of massless fields within which all anomalies happen to cancel. In this sense, the analysis along the lines of the work in Refs.\cite{doran3,doran2} provides a \emph{\/minimal\/} consistent matter spectrum. This is quite familiar from the study of superstring compactification, where it is known that the determination of the exact number of chargeless massless states requires a considerably detailed knowledge about the compactification than is the case with charged matter\cite{polchinski1,becker1,hubsch1}.

 By the same token, and without delving into the dynamics details of the $x_{11}\to0$ limit, we can then specify a \emph{\/minimal\/} consistent massless matter spectrum, to be refined subsequently. Nevertheless, already in its present form, this approach provides results that are both complementary to those in the literature and phenomenologically interesting.

\section{Orbifold Geometry and Anomalies}
\label{OGA}
The compactified space has the structure of a $T^7/\G\simeq(T^6/\ZZ_N)\times(S^1/\ZZ_2)$ orbifold. We take the circle, $S^1$ to be the $x_{11}$ direction and the coordinates of $T^6$ will be specified by $z_1$, $z_2$ and $z_3$ with real coordinates $x_5,\cdots, x_{10}$; for example, we may identify $z_1=x_5+ix_6$, $z_2=x_7+ix_8$, $z_3=x_9+ix_{10}$.

\subsection{Geometry}
 \label{orbimetry}
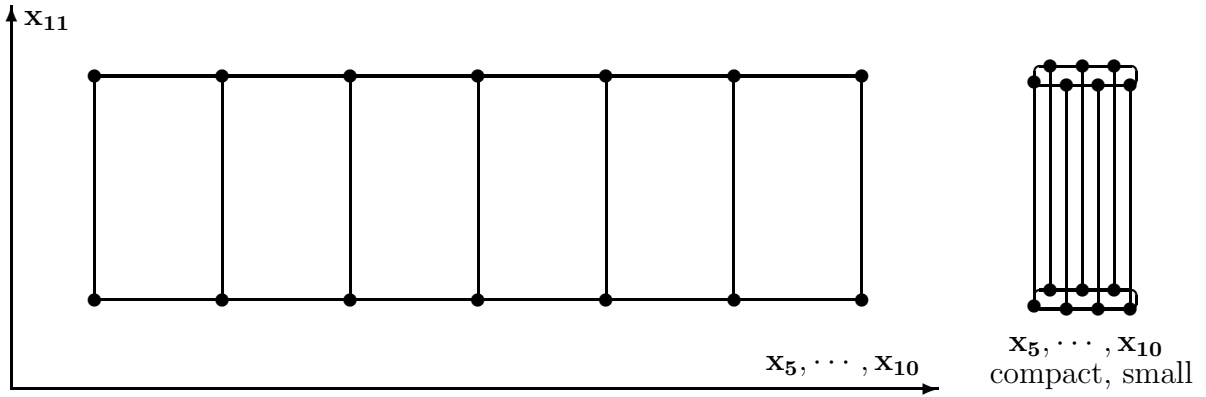
\begin{figure}[ht]
\begin{center}\unitlength=.85mm
\begin{picture}(190, 64) 
\put(2,1){\vector(1,0){145}}
\put(2,1){\vector(0,1){60}}
\put(15,50){\line(1,0){120}}
\put(15,15){\line(1,0){120}}
\put(15,15){\line(0,1){35}}
\put(35,15){\line(0,1){35}}
\put(55,15){\line(0,1){35}}
\put(75,15){\line(0,1){35}}
\put(95,15){\line(0,1){35}}
\put(115,15){\line(0,1){35}}
\put(135,15){\line(0,1){35}}
\put(15,15){\circle*{2}}
\put(35,15){\circle*{2}}
\put(55,15){\circle*{2}}
\put(75,15){\circle*{2}}
\put(95,15){\circle*{2}}
\put(115,15){\circle*{2}}
\put(135,15){\circle*{2}}
\put(15,50){\circle*{2}}
\put(35,50){\circle*{2}}
\put(55,50){\circle*{2}}
\put(75,50){\circle*{2}}
\put(95,50){\circle*{2}}
\put(115,50){\circle*{2}}
\put(135,50){\circle*{2}}
\put(4,58){$\bf{x_{11}}$}
\put(120,4){$\bf{x_5, \cdots , x_{10}}$}
%
%
\put(170,15){\oval(16,3)}
 \put(162,14){\circle*{2}}
 \put(164.5,16.5){\circle*{2}}
 \put(167,13.5){\circle*{2}}
 \put(169.5,16.5){\circle*{2}}
 \put(172,13.5){\circle*{2}}
 \put(174.5,16.5){\circle*{2}}
 \put(177,13.5){\circle*{2}}
\put(170,50){\oval(16,3)}
 \put(162,49){\circle*{2}}
 \put(164.5,51.5){\circle*{2}}
 \put(167,48.5){\circle*{2}}
 \put(169.5,51.5){\circle*{2}}
 \put(172,48.5){\circle*{2}}
 \put(174.5,51.5){\circle*{2}}
 \put(177,48.5){\circle*{2}}
 \put(162,14){\line(0,1){35}}
 \put(164.5,16.5){\line(0,1){35}}
 \put(167,13.5){\line(0,1){35}}
 \put(169.5,16.5){\line(0,1){35}}
 \put(172,13.5){\line(0,1){35}}
 \put(174.5,16.5){\line(0,1){35}}
 \put(177,13.5){\line(0,1){35}}
 \put(158,7){$\bf{x_5, \cdots , x_{10}}$}
 \put(155,2){compact, small}
\end{picture}
\caption[Intersecting planes in the $T^6/\ZZ_7\times S^1/\ZZ_2$ orbifold.]{Intersecting planes in the $T^6/\ZZ_7\times S^1/\ZZ_2$ orbifold: with $x_5,\cdots,x_{10}$ shown unraveled in the sketch on the left-hand side, and curled up on the right-hand side.}
\label{gstr}
\end{center}
\end{figure}

From now on, we focus on $\ZZ_7$;
Table~\ref{acz1} specifies the $\G=\ZZ_7\times\ZZ_2$ quotient group action by listing the fractional multiplets of $2\p i$ phases of the first six coordinates of $T^7$ and the sign of the last coordinate, $x_{11}$. $\ZZ_2$ acts on $x_{11}$ as a parity reflection and leaves invariant two ten-planes at the orbifold singularities; these are shown horizontally in Figure~\ref{gstr}.
\begin{table}[ht]
\begin{center}
\begin{tabular}{ccccccc}
  \multicolumn{1}{c}{\parbox{25mm}{\centering\baselineskip=10pt\bs Orbifold Group}}
 &\multicolumn{1}{c}{\parbox{25mm}{\centering\baselineskip=10pt\bs Group Element}}
 &\multicolumn{1}{c}{\parbox{5mm}{\centering\bs $z_1$}}
 &\multicolumn{1}{c}{\parbox{5mm}{\centering\bs $z_2$}}
 &\multicolumn{1}{c}{\parbox{5mm}{\centering\bs $z_3$}}
 &\multicolumn{1}{c}{\parbox{5mm}{\centering\bs $x_{11}$}}
 &\multicolumn{1}{c}{\parbox{50mm}{\centering\baselineskip=10pt\bs Number and Dimensions
                                                 of ``Hyperplanes''}}\\[2mm]
\hline
&$\a$ & $0$ & $0$ & $0$ & $-$ & 2 ten-planes\\ \cline{2-7}
$\ZZ_7\times\ZZ_2$&$\b$ & $1/7$ & $2/7$ & $-3/7$ & $+$ & 7 five-planes\\ \cline{2-7}
&$\a\b$ & $1/7$ & $2/7$ & $-3/7$ & $-$ & 14 four-planes\\ \hline
\end{tabular}\\[2mm]
\caption{The action of the orbifold group $\ZZ_7\times\ZZ_2$ on the seven compact coordinates. The entries in the middle column are the fractions of $2\p i$ phases; \eg, the entries in the last row signify the action $\a\b:(z_1,z_2,z_3,x_{11})\to(e^{2\p i/7}z_1,e^{4\p i/7}z_2,e^{-6\p i/7}z_3,-x_{11})$.} 
\label{acz1}
\end{center}
\vspace{-2mm}\end{table}
The action of $\ZZ_7$ on the compact coordinates is chosen so as to preserve $N=1$ supersymmetry. It leaves invariant seven five-planes (shown vertically in Figure~\ref{gstr}), spanned by $x_{11}$ and $x_1,\cdots,x_4$, the latter of which are suppressed in Figure~\ref{gstr}. The action of $\ZZ_7\times \ZZ_2$ leaves invariant fourteen four-planes which lie at the intersection of ten-planes and five-planes and represented by the dots in Figure~\ref{gstr}.

In Table~\ref{acz1}, we have omitted the other elements of the orbifold group because their fixed point sets are identical to those of either $\b$ or $\a\b$. Elements $\b^2$, $\b^3$, $\b^4$, $\b^5$ and $\b^6$ are in this sense equivalent to $\b$, and the elements $\a\b^2$, $\a\b^3$, $\a\b^4$, $\a\b^5$ and $\a\b^6$ are equivalent to $\a\b$. The fixed-point sets are called ``hyperplanes'' following the nomenclature in the literature, but we note that such ``$p$-planes'' are in the final orbifold non-compact only in the first four dimensions, $x_1,\cdots,x_4$ (dots in Figure~\ref{gstr}), and compact in the rest: $x_{11}$ is always restricted to the interval $I^1=S^1/\ZZ_2$ (vertical in Figure~\ref{gstr}), and $x_5,\cdots,x_{10}$ are subject to the twisted periodic boundary conditions of $(\IR^6/\L)/\ZZ_7$, as sketched by the ovals in Figure~\ref{gstr}.
%

\subsection{Anomalies}
\label{anom}
On the ten-dimensional plane the chiral projection of the eleven-dimensional supergravity induces gravitational anomaly. To cancel this anomaly we need to introduce Yang-Mills type of gauge fields which further induce gauge and mixed type of anomalies. Presence of some topological terms in the eleven-dimensional supergravity action induce additional anomalies known as `inflow' anomalies. Cancellation of these gravitational, quantum and inflow anomalies require a separate complement of $E_8$ gauge fields in each of the two ten-plane boundary spacetimes\cite{horava1}.

As indicated in Table~\ref{acz1}, within the 11-dimensional bulk and the 10-dimensional boundary, the next $\G$-fixed structure consists of five-planes. This situation is the key difference between our $\ZZ_7$-orbifolds and the models studied earlier\cite{doran3,doran2,doran1}. Therein, each model contained intermediate $\G$-fixed structures between the 10-dimensional and the 5-dimensional one. Those intermediate structures---most notably the $\G$-fixed six-planes---provided pivotal information by way of anomaly cancellation requirements\cite{faux2,faux3,faux4}. Herein, there are no such intermediate structures and therefore no additional constraints enforced by anomaly cancellation in these intermediate $\G$-fixed structures.

So, in the ``next dimension down'', we have five-planes, shown as vertical lines in Figure~\ref{gstr}. One of their five dimensions is spanned by $x_{11}$, and their remaining four dimensions, collapsed to the dots in Figure~\ref{gstr}, lie in each of the two 10-dimensional boundary components of the original $M$-theory.

Since odd-dimensional spacetimes cannot have Weyl fermions, and it is only such fermions that can carry the anomaly, the matter content within the five-planes is anomaly-free.
 However, we will see in the following sections that we must nevertheless introduce some states that extend through the five-planes in order to cancel anomalies on their boundary four-planes.

\section{$\ZZ_7$ Invariant Groups}
\label{z7ig}
On a given four-plane within the boundary ten-plane, the $E_8$ gauge group breaks down to some subgroup ${\cal G}_4$. The breaking depends on the embedding of the $\ZZ_7$ action within the $E_8$ group. The fundamental irreducible representation ${\bf 248}$ of $E_8$, which has rank $8$ is also its adjoint representation. Using {\sl Mathematica}\TM\ programs, we construct all the positive $120$ roots of $E_8$ from the highest root $[0,0,0,0,0,0,1,0]$ and 428 eight-component $\ZZ_7$ vectors with coordinates taking values from among $\{\frac17,\frac27,\frac37,\frac47,\frac57,\frac67\}$, and chosen so as to preserve supersymmetry. Requiring supersymmetry implies that the sum of the components of a vector must add up to an integer\cite{doran1}. Once the roots and the shift vectors are known, the $\ZZ_7$-invariant subgroups of $E_8$ is found by the following procedure.

If $\bm{V}$ denotes a $\ZZ_7$ shift vector acting on the root lattice $\bm{\cal P}$ of group $E_8$, then the set of root vectors that survives the projection
\begin{equation}
 e^{2\p i(\bm{P}\cdot\bm{V})}\ket{\bm{P}} = \ket{\bm{P}},\qquad
 \bm{P}\in\bm{\cal P},
\label{shift}
\end{equation}
constitute the root vectors of a $\ZZ_7$-invariant subgroup $H$ in $E_8$. We denote by $\bm{\cal W}$ the set of 120 positive root vectors of $E_8$.

In the {\bf first} step, we take the dot product between a shift $\bm{V}$ and each of the  roots $\bm{W}_i$'s in $\bm{\cal W}$. A root that satisfies $\bm{V}{\cdot}\bm{W}_i\in\ZZ$ is said to have \emph{\/survived\/} the shift. We denote by $\bm{{\cal T}\!_V}$ this set of surviving roots, and by $k_{\bm{V}}$ the number of surviving roots for a particular shift $\bm{V}$:
\begin{equation}
 k_{\bm{V}}\Defl|\bm{{\cal T}_V}|.
 \label{ekV}
\end{equation}
We consider all the 428 $\ZZ_7$ vectors and their permutations to find out all the possible values of $k_{\bm{V}}$. We extensively use {\sl Mathematica}\TM\ and collect the values of $k_{\bm{V}}$ and find that it can have one of the following values:
\begin{equation}
k_{\bm{V}}\in\{14, 15, 16, 21, 22, 23, 28, 30, 36, 37, 42, 63\}~.
\label{kv}
\end{equation}

In the {\bf second} step, we find all the possible subgroups $H_I\subset E_8$. We use the Dynkin diagram technique and the subgroups are listed in Table~\ref{atsg} in Appendix~\ref{sge8}. We find the number of positive roots $k_H$ for each one of these subgroups. These numbers give us indication of the possible subgroups that \Eq{kv} represents. For example, if the number of surviving roots $k_{\bm{V}}$ for a particular shift $\bm{V}$ is 14, these 14 roots might belong to one of the four subgroups
\begin{equation}
SO_8\times SU_{2}^2,\qquad
SU_5\times SU_3\times SU_2,\qquad
SU_5\times SU_{2}^{4},\quad \text{or}\quad
SU_{4}^2\times SU_{2}^{2},
\end{equation}
as found in Table~\ref{atsg}.

The {\bf third} step is required if the subgroup is not identified unambiguously already by the values of $k_{\bm{V}}$: For each $H_I\subset E_8$, we compute $m$ and $r$ defined as:
\begin{equation}
 \begin{tabular}{rcl}
  $m$ &$\Defl$& the number of $SU(2)$ factors, if any, in $H_I$,\\
  $r$ &$\Defl$& the rank of $H_I$.
 \end{tabular}
 \label{M1}
\end{equation}
Table~\ref{atsg} in Appendix~\ref{sge8} lists the values of $m$ and $r$ used to identify a subgroup of $E_8$.

Finally, in the {\bf fourth} step, we specify a procedure to determine the identifiers $m$ and $r$ from the set of surviving roots found in the first step. This is shown by the following example.

Take the shift vector $\bm{V}=\big[\inv7, \inv7, 0, \frc27, 0, 0, \frc37, 0\big]$, one of the permutations of $\big[\inv7, \inv7, \frc27, \frc37, 0, 0, 0, 0\big]$. The $E_8$ roots that survive the shift are
\begin{equation}
\bm{{\cal T}\!_V} \Defl \left\{~
\begin{matrix}
 [0,0,0,0,1,-1,0,0],& [1,-1,0,0,0,0,0,1],& [1,-1,1,0,0,0,0,-1],\\
 [-1,1,0,0,-1,1,0,0],& [0,0,-1,0,1,0,0,1],& [0,0,0,0,1,0,0,-1],\\
 [1,-1,0,0,0,1,0,0],& [0,0,0,0,0,-1,0,1],& [0,0,1,0,0,-1,0,-1],\\
 [0,1,-1,1,-1,1,-1,0],& [-1,0,1,-1,0,0,1,0],& [1,-1,1,0,-1,0,0,0],\\
 [-1,1,-1,0,1,-1,0,1],& [-1,1,0,0,1,-1,0,-1],& [0,0,-1,0,0,0,0,2]\,
\end{matrix}~\right\}.
 \label{15}
\end{equation}
Let $z^i$, $i=1,2,\cdots15$, denote  these roots. We look for possible relations\ft{A relation of the type $z^i+z^j=z^k$ corresponds to the commutator $[E_{z^i},E_{z^j}]=E_{z^k}$. So, $E_{z^i},E_{z^j}$ and $E_{z^k}$, together with two suitable Cartan generators and $E_{-z^i},E_{-z^j},E_{-z^k}$ generate an $SU(3)$ subgroup of $E_8$. In particular, if  $z^k=z^i+z^j$ for some $z^i$ and $z^j$, then $z^k$ is not a simple root.} in the form $z^i+z^j=z^k$, and find the following:
\begin{subequations}
\label{rrln}
\begin{alignat}{5}
z^2+z^{14}&=z^1~,\quad&\quad
z^3+z^{13}&=z^1~,\quad&\quad
z^3+z^{15}&=z^2~,\\
z^5+z^9&=z^1~,\quad&\quad
z^5+z^{12}&=z^2~,\quad&\quad
z^6+z^8&=z^1~,\\
z^6+z^{12}&=z^3~,\quad&\quad
z^6+z^{15}&=z^5~,\quad&\quad
z^7+z^8&=z^2~,\\
z^7+z^9&=z^3~,\quad&\quad
z^7+z^{13}&=z^5~,\quad&\quad
z^7+z^{14}&=z^6~,\\
z^9+z^{15}&=z^8~,\quad&\quad
z^{10}+z^{11}&=z^4~,\quad&\quad
z^{12}+z^{13}&=z^8~,\\
z^{12}+z^{14}&=z^9~,\quad&\quad
z^{14}+z^{15}&=z^{13}~.
\end{alignat}
\end{subequations}
The roots $z^7,z^{10},z^{11},z^{12},z^{14}$ and $z^{15}$ cannot be expressed as a sum of any other of the roots\eq{15}, which means that they correspond to 6 simple roots, whence the rank of the group must be 6.

Since all 15 roots\eq{15} appear in the system\eq{rrln}, each root is part of a root system larger than $\{-z^i,0,z^i\}$. Therefore, each $SU(2)$ subgroup corresponding to any of the roots\eq{15} is a subgroup of a larger subgroup of $E_8$, and the roots\eq{15} represent no $SU(2)$ factor subgroup of $E_8$. So, looking at the surviving roots we can define the variables $m$ and $r$ as:
\begin{equation}
 \begin{tabular}{rcp{115mm}}
 $m$ &$\Defl$& the number of roots which do not occur in any of the equations of the type $z^i+z^j=z^k$, (in this example, $m=0$)\\
 $r$ &$\Defl$& the number of roots which do not occur on the right-hand side of any of the equations of the type $z^i+z^j=z^k$.
 \end{tabular}
 \label{M2}
\end{equation}
\begin{proof}
The justification for the agreement between the identifications\eq{M1} and\eq{M2} is simple: Note first that $SU(2)$ is the unique Lie group to have the same number, one, of each of a Cartan root, a positive (raising) root and a negative (lowering) root.
 As positive roots that are not themselves sums of other positive roots are by definition {\em\/simple\/}, the one positive root within an $SU(2)$ factor subgroup of $E_8$ clearly must be simple.
 
 Each simple positive root has a corresponding negative and their commutator results in a Cartan generator, thus jointly generating an $SU(2)$ group. In particular, the number of simple roots must then be equal to the rank of the considered subgroup $H\subset E_8$.
 So, simple positive roots that, within a $\bm{{\cal T}\!_V}$, are neither sums of other positive roots nor occur as summands in any of the positive roots, must in fact be separate $SU(2)$ factors within the considered subgroup $H\subset E_8$.
 
In turn, simple positive roots that {\em\/do\/} occur as summands adding up to other positive roots within $\bm{{\cal T}\!_V}$ must belong to bigger root systems, belonging to factors in $H$ that contain $SU(2)$, but are bigger. Partitioning $\bm{{\cal T}\!_V}$ into such root sub-systems decomposes $H$ into individual factors. Within the finite list of subgroups of $E_8$, it is then possible to identify each system of surviving roots $\bm{{\cal T}\!_V}$ with the root system of the corresponding specific subgroup of $E_8$; see Appendix~\ref{nsr}.
\end{proof}

Using Table~\ref{atsg}, this unambiguously identifies the roots\eq{15} as belonging to the group $SO_8\times SU_3$. We employ this analysis in the construction of the {\sl Mathematica}\TM\ codes and find the subgroups of $E_8$ that are invariant under a $\ZZ_7$ shift listed in Table~\ref{giz7s}.

\section{Four-Dimensional Models}
\label{fdm}
There are $14$ possible choices for ${\cal G}_4$ and including the case when $E_8$ is unbroken, there are  $15$ possible groups for the upper-World four-planes and $15$ possible groups for the lower-World four-planes. At this moment there is no restriction on the choice of the upper-World gauge group and lower-World gauge group, so in the limit  $x_{11}\rightarrow 0$, when the two worlds fuse (for comparison with heterotic string compactification), we end up getting ${15+1\choose2}=120$ possible groups on a four-plane. We look at one of these possibilities and investigate the four-dimensional physics. Take for example $E_8\rightarrow SU_5\times SU_3\times SU_2\times U_1$ breaking. The branching\ft{The double-headed arrow, $\TO$, denotes a projection to a subset.} rule is\cite{slansky1}:
\begin{eqnarray}
E_8 &\rightarrow &SU_5\times SU_5\rightarrow SU_5\times SU_3\times SU_2\times U_1:\nn \\
{\bf 248} &=& ({\bf 1,24})\op ({\bf 24,1})\op ({\bf 5,\ba {10}})\op ({\bf \bar {5},10})\op ({\bf 10,5})\op ({\bf \ba {10},\bar {5}})\nn \\
&=& [({\bf 1,1,1})_{\cc}\op ({\bf 1,1,3})_{\cc}\op ({\bf 1,3,2})_{(-5)}\op ({\bf 1,\bar{3},2})_{(5)}\op ({\bf 1,8,1})_{\cc}]\op ({\bf 24,1,1})_{\cc}\nn\\
&& \op[({\bf 5,1,1})_{(-6)}\op ({\bf 5,3,1})_{(4)}\op ({\bf 5,\bar{3},2})_{(-1)}]\op [({\bf \bar {5},1,1})_{(6)}\op ({\bf \bar {5},\bar{3},1})_{(-4)}\op ({\bf \bar {5},3,2})_{(1)}]\nn\\
&&\op[({\bf 10,1,2})_{(3)}\op ({\bf 10,3,1})_{(-2)}]\op [({\bf \ba {10},1,2})_{(-3)}\op ({\bf \ba {10},\bar{3},1})_{(2)}]\nn\\
&\TO&({\bf 1,1,1})_{\cc}\op ({\bf 1,1,3})_{\cc}\op ({\bf 1,3,2})_{(-5)}\op({\bf 1,8,1})_{\cc}\op ({\bf 24,1,1})_{\cc}\op ({\bf 5,1,1})_{(-6)}\nn\\
&&\op({\bf 5,\bar{3},2})_{(-1)}\op ({\bf 5,3,1})_{(4)}\op ({\bf 10,3,1})_{(-2)}\op ({\bf 10,1,2})_{(3)}~.
\end{eqnarray}
On a four plane only the vector multiplets and the chiral multiplets survives, so at the end of the branching we keep only the ${\bf\cal R}$ part for the representations of the form ${\bf\cal R}\op {\bf\bar{\cal R}}$. The part of the surviving representations that contribute to the anomaly is
\begin{eqnarray}
&({\bf 1,3,2})_{(-5)}\op ({\bf 5,1,1})_{(-6)}\op ({\bf 5,\bar{3},2})_{(-1)}\op ({\bf 5,3,1})_{(4)}\op ({\bf 10,3,1})_{(-2)}\op ({\bf 10,1,2})_{(3)}~.
\end{eqnarray}
To cancel the anomaly each of the following terms must vanish,
\begin{eqnarray}
I({\cal G}_{I}) &\equiv&\sum_i\,\sum_{{\bf {\cal R}}_I}\, n({\bf {\cal R}}_I)_i\,\,q_i\, I_2({\bf {\cal R}}_I)~,\nn\\
 I(U_1)_{{\rm GAUGE}} &=& \sum_i\,N_i\,q_i^3~.
\end{eqnarray}
where ${\cal G}_{I}$ are each of the group factor, $n({\bf {\cal R}}_I)_i$ are the multiplicity of each of the representation ${\bf {\cal R}}_I$ in the $i$-th set of fields, $q_i$ is the charge of each set of fields, $I_2({\bf {\cal R}}_I)$ is the second index of the representation  ${\bf {\cal R}}_I$ and $N_i$ is the total number of fields in each set. Note that all of these multiplets have a factor of $1\over f$ from the index theorem, where $f$ is number of five-planes for the corresponding $\ZZ_N$ orbifold. On a given four-plane the total anomaly turns out to be
\begin{equation}
\begin{aligned}
I(SU_5) &= \frac{1}{f}\left(1(-6)I_2({\bf 5})+6(-1)I_2({\bf 5})+3(4)I_2({\bf 5})+3(-2)I_2({\bf 10})+2(3)I_2({\bf 10})\right)\\
&= \frac{1}{f}\left(1(-6){\cdot}1+6(-1){\cdot}1+3(4){\cdot}1+3(-2){\cdot}3+2(3){\cdot}3\right)~=~0~.\\
I(SU_3) &=  \frac{1}{f}\left(2(-5)I_2({\bf 3})+10(-1)I_2({\bf \bar{3}})+5(4)I_2({\bf 3})+10(-2)I_2({\bf 3})\right)\\
&=  \frac{1}{f}\left(2(-5){\cdot}1+10(-1){\cdot}1+5(4){\cdot}1+10(-2){\cdot}1\right) \,=\, \frac{1}{f}(-20)~\neq~0~.\\
I(SU_2) &=  \frac{1}{f}\left(3(-5)I_2({\bf 2})+15(-1)I_2({\bf 2})+10(3)I_2({\bf 2})\right)\\
&=  \frac{1}{f}\left(3(-5){\cdot}1+15(-1){\cdot}1+10(3){\cdot}1\right)\,=\,0~.\\
I(U_1)_\text{GAUGE}&=\frac{1}{f}\left(6(-5)^3+5(-6)^3+30(-1)^3+15(4)^3+30(-2)^3+20(3)^3\right)
\neq 0~.
\end{aligned}
\end{equation}

So there are gauge and $U_1$ anomalies for this example. Of all the possible choices of gauge groups in Table~\ref{giz7s}, only three of them do not contribute to any anomaly. They are $E_7\times U_1$, $SO_{12}\times U_{1}^{2}$ and $SO_{10}\times SU_2\times U_{1}^{2}$. The branching rules for these three groups are,

\noindent 1. $E_8\rightarrow E_7\times SU_2\rightarrow E_7\times U_1$ :
\begin{eqnarray}
{\bf 248} &=& ({\bf 1,3})\oplus ({\bf 133,1})\oplus ({\bf 56,2})\nonumber \\
&=& \big[{\bf 1}_{(-1)}\oplus{\bf 1}_{(0)}\oplus{\bf 1}_{(1)}\big]\oplus {\bf 133}_{(0)}\oplus\big[{\bf 56}_{(-{1\over 2})}\oplus{\bf 56}_{({1\over 2})}\big]~.
\end{eqnarray}
2. $E_8\rightarrow SO_{16}\rightarrow SO_{12}\times SU_2\times SU_2\rightarrow SO_{12}\times U_{1}^{2}$ :
\begin{eqnarray}
{\bf 248} &=& {\bf 120}\oplus {\bf 128}\nonumber \\
&=& \big[({\bf 66,1,1})\oplus ({\bf 12,2,2})\oplus ({\bf 1,1,3})\oplus ({\bf 1,3,1})\big]\oplus \big[({\bf 32,2,1})\oplus ({\bf 32,1,2})\big]\nonumber \\
&=& \Big[({\bf 66,1})_{(0)}\oplus [({\bf 12,2})_{(-{1\over 2})}\oplus ({\bf 12,2})_{({1\over 2})}\big]\oplus \big[({\bf 1,1})_{(-1)}\oplus ({\bf 1,1})_{(0)}\oplus ({\bf 1,1})_{(1)}\big]\oplus ({\bf 1,3})_{(0)}\Big]\nn\\
&& \op\Big[({\bf 32,2})_{(0)}\oplus \big[({\bf 32,1})_{(-{1\over 2})}\oplus ({\bf 32,1})_{({1\over2})}\big]\Big]\nonumber \\
&=& \Big[{\bf 66}_{(0)(0)}\oplus\big[[{\bf 12}_{(-{1\over 2})(-{1\over 2})}\oplus {\bf 12}_{({1\over 2})(-{1\over 2})}]\oplus [{\bf 12}_{(-{1\over 2})({1\over 2})}\oplus {\bf 12}_{({1\over 2})({1\over 2})}]\big]\oplus [{\bf 1}_{(0)(-1)}\oplus {\bf 1}_{(0)(0)}\oplus {\bf 1}_{(0)(1)}]\nonumber \\
&& \oplus [{\bf 1}_{(-1)(0)}\oplus {\bf 1}_{(0)(0)}\oplus {\bf 1}_{(1)(0)}]\Big]\oplus \Big[[{\bf 32}_{(-{1\over 2})(0)}\oplus {\bf 32}_{({1\over 2})(0)}]\oplus [{\bf 32}_{(0)(-{1\over 2})}\oplus {\bf 32}_{(0)({1\over2})}]\Big]~.
\end{eqnarray}
3. $E_8\rightarrow SO_{16}\rightarrow SO_{10}\times SU_4\rightarrow SO_{10}\times SU_{2}^{2}\times U_1\rightarrow SO_{10}\times SU_2\times U_{1}^{2}$ :
\begin{eqnarray}
{\bf 248} &=& {\bf 120}\oplus {\bf 128}\nonumber \\
&=& [({\bf 45,1})\oplus ({\bf 10,6})\oplus ({\bf 1,15})]\oplus [({\bf 16,4})\oplus ({\bf 16,\bar 4})]\nonumber \\
&=& \big[({\bf 45,1,1})_{(0)}\oplus [({\bf 10,1,1})_{(2)}\oplus ({\bf 10,1,1})_{(-2)}\oplus ({\bf 10,2,2})_{(0)}]\nonumber \\
&& \oplus [({\bf 1,1,1})_{(0)}\oplus ({\bf 1,3,1})_{(0)}\oplus ({\bf 1,1,3})_{(0)}\oplus ({\bf 1,2,2})_{(2)}\oplus ({\bf 1,2,2})_{(-2)}]\big]\nonumber \\
&&\oplus \big[[({\bf 16,2,1})_{(1)}\oplus ({\bf 16,1,2})_{(-1)}]\oplus [({\bf 16,2,1})_{(-1)}\oplus ({\bf 16,1,2})_{(1)}]\big]\nonumber \\
&=& \big[({\bf 45,1})_{(0)(0)}\oplus [({\bf 10,1})_{(0)(2)}\oplus ({\bf 10,1})_{(0)(-2)}\oplus [({\bf 10,2})_{(-{1\over 2})(0)}\oplus ({\bf 10,2})_{({1\over 2})(0)}]\big]\nonumber \\
&& \oplus \Big[({\bf 1,1})_{(0)(0)}\oplus ({\bf 1,3})_{(0)(0)}\oplus [({\bf 1,1})_{(-1)(0)}\oplus ({\bf 1,1})_{(0)(0)}\oplus ({\bf 1,1})_{(1)(0)}]\nn\\
&&\op[({\bf 1,2})_{(-{1\over 2})(2)}
\oplus ({\bf 1,2})_{({1\over 2})(2)}]\oplus [({\bf 1,2})_{(-{1\over 2})(-2)}\oplus ({\bf 1,2})_{({1\over 2})(-2)}]\big]\Big]\nn\\
&&\oplus \Big[\big[({\bf 16,2})_{(0)(1)}\oplus [({\bf 16,1})_{(-{1\over 2})(-1)}\oplus ({\bf 16,1})_{({1\over 2})(-1)}]\big]\nn\\
&&\oplus \big[({\bf 16,2})_{(0)(-1)}\oplus [({\bf 16,1})_{(-{1\over 2})(1)}\oplus ({\bf 16,1})_{({1\over 2})(1)}]\big]\Big]~.
\end{eqnarray}
However, the final massless contents of these choices are non-chiral. So, we look for a possible solution to this problem. So far, we don't have any mechanism to cancel anomaly on a given four-dimensional plane, which requires twisted states and possibly some contribution from the five-planes. We now turn to these.

\section{$M$-Theory Models from String Theory Orbifolds}
\label{msor}
We explained in the introduction how four-dimensional orbifold models constructed from $E_8\times E_8$ heterotic string theory are related to those constructed in $M$-theory. In this chapter we would like to emphasize the relationship a little further. In order to do that, we need to consider the possible four-dimensional string compactifications on ``our'' orbifolds. In\cite{katsuki1,katsuki2}, Katsuki, {\it et al.}  have listed all the possible gauge groups on a four-dimensional plane for $\ZZ_N$ orbifolds in general. For $\ZZ_7$ orbifold there are 39 such models. They have also listed massless twisted and untwisted states for all the $\ZZ_N$ orbifold models, except for $\ZZ_8$ and $\ZZ_{12}$ because there are too many massless states in those cases, and this number increases as one goes to higher order orbifolds. However the $U(1)$ charges of the matter representations were omitted. These $U(1)$ charges are necessary if one wants to relate the model with one in the $M$-theory construction. We follow\cite{dine1,kobayashi1} to calculate these charges.

\subsection{Anomalous $U(1)$ Gauge Symmetry}
\label{augs}
All the models listed in\cite{katsuki1,katsuki2} have at least one $U(1)$ factor. As we will show, in any of these models we can chose the $U(1)$ basis in such way that at most one $U(1)$ factor is anomalous. The presence of $U(1)$ anomalies are allowed in this kind of orbifold theories as they can be cancelled by a four-dimensional version of the universal Green-Schwarz mechanism\cite{green1}.

Lets take the anomalous gauge group, $\prod_a G_a\times U(1)_A$. The charge $Q_A$ associated with the anomalous $U(1)$ factor is related to a constant $\d_\text{GS}$ by $\Tr Q_A=96\p^2\sqrt{k_A}\d_\text{GS}$. Here $k_A$ is the level of $U(1)_A$ and the constant $\d_\text{GS}$ has a unique value. For models with only one $U(1)$ factor the following universality relation is satisfied,
\begin{equation}
{1\over k_a}\,\Tr_{G_a}\,T(R)\,Q_A={1\over 3}\,\Tr \,Q_{A}^{3}={1\over 24}\,\Tr \,Q_A\equiv 8\p^2\d_\text{GS}~,
\label{gs}
\end{equation}
where $2T(R)$ is the index of the representation $R$ and the charge $Q_A$ has been rescaled so that $k_A=1$. This is referred to as the universal GS relation. By forming suitable linear combinations of more than one $U(1)$ factor, it is always possible to change the basis so that at most one $U(1)$ factor is anomalous. The universal relation in that case is,
\begin{equation}
{1\over k_a}\,\Tr_{G_a}\,T(R)\,Q_A=\Tr \,Q_{B}^{2}\,Q_A={1\over 3}\,\Tr \,Q_{A}^{3}={1\over 24}\,\Tr \,Q_A=8\p^2\d_\text{GS}~,
\end{equation}
where $Q_B$ is the charge with respect to any of the non-anomalous $U(1)$ factors.

To illustrate the above procedure we give an example: Consider the model, $E_8\times E'_8\rightarrow SU(8)\times U(1)_A\times E_6\times SU(2)\times U(1)_B$. We use {\sl Mathematica}\TM\ to find out the untwisted and twisted states along with their charges. 

\begin{enumerate}\itemsep=-3pt\vspace{-3mm}
 \item[] {\bf\underline {Untwisted States}:}\\*[-5mm]
 \item[{\it i.}] We find out the 248 roots of $E_8$ from the following.
\begin{equation}
 P^I=[\underline{\pm1, \pm1, 0,0,0,0,0,0}]\equiv [\underline{\pm1, \pm1, 0^6}]~,
 \label{roots1}
\end{equation}
\begin{equation}
 P^I=[\pm \inv2,\pm \inv2,\pm \inv2,\pm \inv2,\pm \inv2,\pm \inv2,\pm \inv2,\pm \inv2]~,
 \label{roots2}
\end{equation}
where underline indicates the inclusion of all permutations, and one must select an even number of minus signs in\eq{roots2}. We also include the 8 Cartan roots, $[0,0,0,0,0,0,0,0]$.
 \item[{\it ii.}] Multiply these roots $P^I$ with the shift $V^I$. For the example at hand, $V^I=\inv7[1,1,1,1,1,1,1,-1]$ for the group $SU(8)\times U(1)_A$. Look for roots that give $P^I{\cdot}V^I\in\ZZ$. This gives us the untwisted states.
 \item[{\it iii.}] Look for a suitable U(1) basis. The general procedure is described below\ft{Thanks to Tatsuo Kobayashi for explaining this.}:
\end{enumerate}\vspace{-3mm}
\begin{quotation}\vspace{-5mm}
\noindent{\it A Suitable $U(1)$ Basis:}\\*
The $E_8$ Kac-Moody algebra consists of the $\partial X^I$ and $e^{iP^IX^I}$ operators, where $P^I$ are nonzero roots of $E_8$ and $\partial X^I$ corresponds to Cartan parts. Now, the shift vector $V^I$ breaks the $E_8$ group to a gauge group, whose Kac-Moody algebra consists of $\partial X^I$ and $e^{iP^IX^I}$, where $P^I$ are the surviving roots satisfying $P^I{\cdot}V^I\in\ZZ$. Suppose that the unbroken gauge group is $G_7 \times U(1)$, for some simple rank-7 subgroup $G_7\subset E_8$. This $U(1)$ corresponds to a linear combination of $\partial X^I$, \ie, $\sum_I a_I \partial X^I$. These coefficients are fixed so that $\sum_I a_I \partial X^I$ commutes with the non-abelian part, $G_7$, implying that 
\begin{equation}
\sum_I a_I P^I =0~.
\end{equation}
The procedure to fix a U(1) basis is essentially the same even when the unbroken gauge group includes two or more $U(1)$ factors. For an example, let us consider the case where the unbroken gauge group is $G_6 \times U(1)^2$. The two $U(1)$ factors correspond to linear combinations of $\partial X^I$, \ie, $\sum_I a^{(1)}_I \partial X^I$ and $\sum_I a^{(2)}_I \partial X^I$. They should satisfy 
\begin{equation}
\sum_I a^{(1)}_I P^I =0, \qquad \sum_I a^{(2)}_I P^I =0~, 
\end{equation}
for $P^I$ satisfying $P^I{\cdot}V^I\in\ZZ$, and they should be orthogonal to each other, \ie, 
\begin{equation}
\sum_I a^{(1)}_I a^{(2)}_I =0~.
\end{equation}
In the $(a^{(1)},a^{(2)})$-plane, there exists one degree of freedom to rotate the basis. If both $U(1)$ factors are anomaly-free, we can choose any basis. If one of the $U(1)$ factors is anomalous, we identify it. Finally, if both $U(1)$ factors are anomalous, we rotate to a basis wherein one $U(1)$ is anomaly-free, leaving the other to be anomalous.
\end{quotation}

In this way, we find the $U(1)$ basis for the shift $V^I=\inv7[1,1,1,1,1,1,1,-1]$ is given by $7V^I$. Once the basis vector is fixed, the product of the basis vector with the surviving $P^I$ gives the untwisted states and their charges as:
\begin{equation}
{\bf(56)}_{-1}\oplus {\bf (\ba{28})}_{-2}\oplus {\bf(\bar{8})}_{3}~.
\end{equation}
We do the same for the other $E'_8$ factor, where the shift is $V^I=\inv7[1,1,1,1,1,1,0,0]$, the $U(1)$ charge basis is again $7V^I$, and the gauge group is now $E_6\times SU(2)\times U(1)_B$. We get the untwisted states:
\begin{equation}
{\bf (27,2)}_{1}\oplus{\bf(\ba{27},1)}_{2}\oplus {\bf(1,2)}_{-3}~.
\end{equation}
Together, the $SU(8)\times U(1)_A\times E_6\times SU(2)\times U(1)_B$ representation of the untwisted states is
\begin{equation}
{\bf(56,1,1)}_{-1,0}\oplus {\bf (\ba{28},1,1)}_{-2,0}\oplus {\bf(\bar{8},1,1)}_{3,0}\oplus {\bf (1,27,2)}_{0,1}\oplus{\bf(1,\ba{27},1)}_{0,2}\oplus {\bf(1,1,2)}_{0,-3}~.
\label{unts}
\end{equation}

\goodbreak
\begin{enumerate}\itemsep=-3pt\vspace{-3mm}
 \item[] {\bf\underline {Twisted States}:}\\*[-5mm]
 \item[{\it i.}] Find the 248 roots of $E_8$ from \Eq{roots1} and \Eq{roots2}.
 \item[{\it ii.}] Multiply the shift $V^I$ with $k$, the twisted sector index. If any of the components of $k\,V^I$ is outside $[-3/7,3/7]$, we simplify as follows: Take for example $V^I=\inv7[3,3,2,2,2,2,1,-1]$ and $k=4$, this gives $k\,V^I=\inv7[12,12,8,8,8,8,4,-4]$. Now we simplify $k\,V^I$ by reducing each component of $7(k\,V^I)$ modulo 7, producing $k\,V^I\cong\inv7[-2,-2,1,1,1,1,-3,3\}\Defr w$.
 \item[{\it iii.}] Calculate the shifted momenta $\Tw{P}^I= P^I+w$.
 \item[{\it iv.}] Do the same for the other $E'_8$ factor. Call this $\Tw{P}^{I'}$.
 \item[{\it v.}] Compute the momenta $\tilde{P}^{\hat I}=(\tilde{P}^I,\tilde{P}^{I'})$ and then $\sum_{{\hat I}=1}^{16}(\tilde{P}^{\hat I})^2$. This gives a total of $248\times 248$ values.
 \item[{\it vi.}] Calculate $\sum_{{\hat I}=1}^{16}(\tilde{P}^{\hat I})^2$ from the mass formula for the left moving twisted sectors (in the light-cone gauge),
\begin{equation}
{1\over 8}m_{L}^{2}={1\over 2}\sum_{i=1}^{3}\left(p_{L}^{i}\right)^2\d_{k,0}+{1\over 2}\sum_{\hat{I}=1}^{16}\left(\tilde{P}^{\hat{I}}\right)^2+N_{L}^{(k)}-1+c^{(k)}~.
\label{ml2}
\end{equation}
The term $N_{L}^{(k)}$ is the oscillator number and takes the fractional value which is  multiple of ${1\over N}$, where $N$ is the order of the orbifold group. For $\ZZ_7$ orbifold the zero-point energy is same for any twisted sector and is equal to $c^{(k)}=2/7$. The untwisted states corresponds to $k=0$ and the momenta $p_{L}^i$ are associated with the compact coordinates $X_{L}^j$, where $j=3,\cdots,8$. For twisted states the term $\sum(p_{L}^{i})^2$ does not contribute to the mass equation. Since the value of $N_{L}^{(k)}$ is a multiple of $\frac{1}{N}$, we get $N_{L}^{(k)}= 0,\frac{1}{7},...\,,\frac{6}{7}$. So, from\eq{ml2} we get
\begin{equation}
\sum_{\hat{I}=1}^{16}(\tilde{P}^{\hat{I}})^2= -2N_{L}^{(k)}+\frac{10}{7}~.
\end{equation}
 For the various possible values of $N_{L}^{(k)}$,
\begin{equation}
\sum_{\hat{I}=1}^{16}(\tilde{P}^{\hat{I}})^2=\frac{10}{7}\,,\frac{8}{7}\,,\frac{6}{7}\,,\frac{4}{7}\,,\frac{2}{7}\,,0\,,-\frac{2}{7}~.
\end{equation}
Since $\sum_{\hat{I}=1}^{16}(\tilde{P}^{\hat{I}})^2$ is positive, the maximum and minimum values of  $\sum_{\hat{I}=1}^{16}(\tilde{P}^{\hat{I}})^2$ are $\frac{10}{7}$ and $0$, respectively.
This gives $\sum_{{\hat I}=1}^{16}(\tilde{P}^{\hat I})^2={10\over 7}, {8\over 7}, {6\over 7}, {4\over 7}, {2\over 7} $ for $7N_{L}^{(k)}=0,1,2,3,4$ respectively.
 \item[{\it vii.}] Now look for $\tilde{P}^{\hat I}$ in step {\it v.}, which gives $\sum_{{\hat I}=1}^{16}(\tilde{P}^{\hat I})^2={10\over 7}, {8\over 7}, {6\over 7}, {4\over 7}$ or ${2\over 7}$.
 \item[{\it viii.}] Take the same U(1) basis as in the untwisted case and multiply it with the surviving $\tilde{P}^{\hat I}$. This produces the twisted states and their charges in the $k=1,2\text{ and }4$ sectors as follows, 
\begin{eqnarray}
k=1&:&\quad 7{\bf(\bar{8};1,1)}_{-{13\over 7},{6\over 7}}\oplus 7{\bf (1;1,2)}_{{8\over 7},-{15\over 7}}\oplus 35{\bf (1;1,1)}_{{8\over 7},{6\over 7}}~,\nn\\
k=2&:&\quad 7{\bf(\bar{8};1,1)}_{-{5\over 7},{12\over 7}}\oplus7{\bf(1;1,2)}_{{16\over 7},-{9\over 7}}\oplus 7{\bf (1;1,1)}_{{16\over 7},{12\over 7}}~,\nn\\
k=4&:&\quad 7{\bf(8;1,2)}_{-{3\over 7},{3\over 7}}~.
\label{twis23}
\end{eqnarray}
\end{enumerate}\vspace{-3mm}

\noindent{\bf\underline {Anomaly}:}\nobreak\vglue0mm\noindent
For the group $SU(8)\times U(1)_A\times E_6\times SU(2)\times U(1)_B$ the shift may be written as $V^{\hat{I}}=(1^7,-1)(1^6,0^2)/7$. We have two charge basis vectors: $V_{A}^{\hat {I}}=(1^7,-1)(0^8)$ and $V_{B}^{\hat {I}}=(0^8)(1^6,0^2)$. From the charge basis vectors, we calculate the level of $U(1)$ as
\begin{equation}
k_Q=2\sum_{\hat{I}=1}^{16}(V_{Q}^{\hat{I}})^2~.
\end{equation}
This gives $k_A=16$ and $k_B=12$. Note that there is no $E_6$ representation in the twisted sector, which tells us that $U(1)_B$ is anomalous. Using the representation index in\cite{mckay1,slansky1} we calculate the $\Tr_{G_a}\,T(R)\,Q_B$, $\Tr Q_B$, $\Tr Q_{B}^{3}$ and $\Tr Q_{A}^{2}Q_B$ as
\begin{eqnarray}
\Tr_{SU_8}T(R)Q_B&=&7\times {1\over 2}\times {6\over 7}+7\times {1\over 2}\times {12\over 7}+14\times {1\over 2}\times {3\over 7}=12~,\nn\\
\Tr_{E_6}T(R)Q_B&=&2\times {6\over 2}\times 1+1\times {6\over 2}\times 2=12~,\nn\\
\Tr_{SU_2}T(R)Q_B&=&27\times {1\over 2}\times 1+1\times {1\over 2}\times (-3)+7\times {1\over 2}\times \left({-15\over 7}\right)\nn\\
&&+7\times {1\over 2}\times \left({-9\over 7}\right)+56\times {1\over 2}\times {3\over 7}=12~,\nn\\
\Tr Q_B&=&54+54-6+186=288~,\nn\\
\Tr Q_{B}^{3}&=&54+216-54+216=432~,\nn\\
\Tr Q_{A}^{2}Q_B&=&192~.
\end{eqnarray}
With these values, all the traces are related as
\begin{equation}
\Tr_{G_a}T(R)Q_B={1\over 16}\Tr Q_{A}^{2}Q_B={1\over 36}\Tr Q_{B}^{3}={1\over 24}\Tr Q_B=12~.
\end{equation}
We rescale the charge as $Q_A\rightarrow \sqrt{k_A}\,Q_A$ and $Q_B\rightarrow \sqrt{k_B}\,Q_B$ which gives the universal GS relation
\begin{equation}
\Tr_{G_a}T(R)Q_B=\Tr Q_{A}^{2}Q_B={1\over 3}\Tr Q_{B}^{3}={1\over 24}\Tr Q_B=12~.
\end{equation}
In this way it is possible to find all the charges of the twisted and untwisted matter content in\cite{katsuki1} and to calculate the universal GS relation, ensuring all models to be consistent and anomaly free.

\subsection{Four-Dimensional $M$-Theory Models}
\label{fdmm}
In Section~\ref{fdm}, we found ${15+1\choose2}$ $M$-theory models. But all of them were either anomalous or free of chiral fermions. In analogy with the string theory models, now we conclude that there are 39 four-dimensional $M$-theory models which are free of anomalies. Gratifyingly, they turn out to be the same models as are listed in\cite{katsuki1,katsuki2}. If one of the gauge groups in Table-6 of\cite{katsuki1} represents the symmetry group of the upper four-plane (Figure~\ref{gstr}) then the other one represents the group on the lower four-plane. In the limit $x_{11}\rightarrow 0$, the matter spectrum for a model is given by the untwisted and twisted states in Table-6 and Table-23 of\cite{katsuki1}.

In the $SU(8)\times U(1)_A\times E_6\times SU(2)\times U(1)_B$ example given in the previous section, the untwisted states are represented by \Eq{unts} and the twisted states are represented by \Eq{twis23}. This is an interesting model,  in the $k=4$ twisted sector the states are charged with respect to the gauge groups of both the lower and the upper 4-dimensional World, \ie, both $SU_8$ and $E_6\times SU_2$, respectively.

Such states, which carry charges in both 4-dimensional Worlds, {\em must\/} be represented by an extended object, which stretches from one World to the other. In this case, it means that the seven $({\bf8,1,2})$-states must be representable by an extended object (a suitable $p$-brane) stretching along the $x_{11}$ coordinate, with one ``end'' constrained to one World, the other to the other World. By the same argument, {\em any\/} state which must carry nonzero gauge-charges with respect to the gauge groups in both the ``lower'' and the ``upper'' World, must be represented by an extended object, stretched along $x_{11}$ from one World to the other. In effect, such models include modes of interaction between two otherwise isolated four-dimensional Worlds, separated by the $x_{11}$ coordinate.

From Table-23 of\cite{katsuki1}, we find that there are 10 models where this phenomenon occurs. In our $M$-theory models, it is the five-planes that connect the two four-plane Worlds, so these twisted states must ``live'' in the five-planes. This is the contribution from the five-planes that we were expecting in Section~\ref{anom}.

\section{Four-Dimensional Gauge Groups with One Wilson Line}
\label{4dggwl}
If we denote the lattice of the six-dimensional torus by $\L$, then for $\ZZ_7$ orbifold $\L$ must be the $SU_7$ lattice, which in turn implies that we can add only one Wilson line. For $\ZZ_8$ and $\ZZ_{12}$ the number of Wilson lines depends on the number of lattice choices for $\L$, which can be found in\cite{ibanez1}. These Wilson lines provide extra shifts and break the four-dimensional gauge groups to smaller and more realistic groups such as $SU_3\times SU_2\times U_{1}^{5}$.

In Section~\ref{z7ig} we have shown how to find a subgroup of $E_8$ that survives a $\ZZ_7$ shift vector $V$. There is an alternate way of identifying the surviving gauge group\cite{choi1}. This is known as Dynkin diagram technique, originally proposed by Kac and Peterson\cite{kac1}. In this technique, one takes the extended Dynkin diagram $E_8$ and looks for a redefined shift vector, which satisfies the same condition as before, $P{\cdot}V\in\ZZ$. This results in removing some circles from the Dynkin diagram which gives the surviving subgroup. Herein, we follow the work of\cite{choi1}.

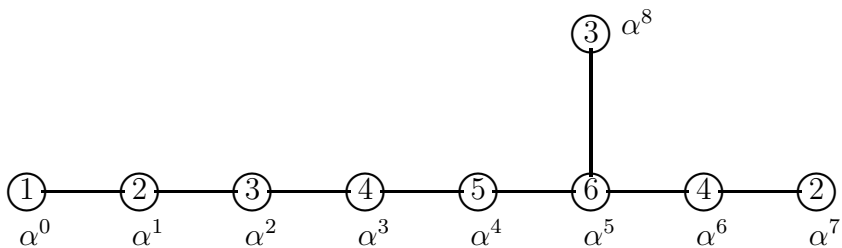
\begin{figure}[h]
\begin{center}
\begin{picture}(100,25)
\put(0,0){\circle{5}}
\put(15,0){\circle{5}}
\put(30,0){\circle{5}}
\put(45,0){\circle{5}}
\put(60,0){\circle{5}}
\put(75,0){\circle{5}}
\put(90,0){\circle{5}}
\put(105,0){\circle{5}}
\put(75,21){\circle{5}}
\put(2,0){\line(1,0){11}}
\put(17,0){\line(1,0){11}}
\put(32,0){\line(1,0){11}}
\put(47,0){\line(1,0){11}}
\put(62,0){\line(1,0){11}}
\put(77,0){\line(1,0){11}}
\put(92,0){\line(1,0){11}}
\put(75,2){\line(0,1){17}}
\put(-1,-1){$1$}
\put(14,-1){$2$}
\put(29,-1){$3$}
\put(44,-1){$4$}
\put(59,-1){$5$}
\put(74,-1){$6$}
\put(89,-1){$4$}
\put(104,-1){$2$}
\put(74,20){$3$}
\put(-1,-7){$\a^0$}
\put(14,-7){$\a^1$}
\put(29,-7){$\a^2$}
\put(44,-7){$\a^3$}
\put(59,-7){$\a^4$}
\put(74,-7){$\a^5$}
\put(89,-7){$\a^6$}
\put(104,-7){$\a^7$}
\put(79,21){$\a^8$}
\end{picture}
\vspace{5mm}
\caption[Extended Dynkin diagram of $E_8$]{Extended Dynkin diagram of $E_8$. The numbers in the circles are the Kac labels. The extended root is $\a^0$.}
\label{e8d}
\end{center}
\end{figure}

The  extended Dynkin diagram of $E_8$ is shown in Fig.~\ref{e8d}. The highest root $\q$ of $E_8$ is given by
\begin{equation}
\q=\sum_{i=1}^{r}n_i\a^i=2\a^1+3\a^2+4\a^3+5\a^4+6\a^5+4\a^6+2\a^7+3\a^8\,,
\end{equation}
where $\a^i$, $i=1,\cdots,8$, are the simple roots of $E_8$, $r$ is the rank and the coefficients $n_i$ are are the Kac labels. In the diagram, $\a^0$ is the negative of the highest root, $\a^0=-\q$.
We expand the $\ZZ_7$ shift vector in the Dynkin basis $\{\g_i\}$, satisfying $\g_i{\cdot}\a^j=\d_{i}^{j}$,
\begin{equation}
V={1\over N}\sum_{i=1}^{r}s_i\g_i~,
\end{equation}
where $N$ is the order of the shift vector and $s_i=N\,\a^i{\cdot}V$. We define $s_0$ as
\begin{equation}
s_0\equiv N(1-\q{\cdot}V)=N-\sum_{i=1}^{8}n_is_i~,
\end{equation}
and define a Kac label for the simple root $\a^0$ as $n_0=1$. We look for a set of integers $s_I$, where $I = 0,...,8$, that satisfy
\begin{equation}
\sum_{I=0}^{8}n_Is_I=N,\qquad s_I\geq0~.
\label{nisi}
\end{equation}
For each nonzero value in the set $s_I$, we remove the $I$-th circle from the extended Dynkin diagram and read off the surviving group from the remaining circles in the diagram. Let $G=G_1\times G_2\times\cdots$ be the unbroken subgroup obtained in this way, where each $G_x$ factor is a simple group.

Our next step is to add a Wilson line which can be realized by a shift vector `$a$'. We expand `$a$' in terms of the $E_8$'s fundamental weights as
\begin{equation}
a={1\over N}\sum_{i=1}^{8}w_i\g_i~,
\end{equation}
where the set of integers $w_i\geq0$ satisfy a similar condition as $s_I$ in \Eq{nisi}. We define the highest root $\q^{G_x}=-\a^{G_x}$ of the simple group $G_x$ as
\begin{equation}
\q^{G_x}=\sum_{i\in J_x}n_{i}^{G_x}\a^i~,
\end{equation}
where $J_x$ is the set of indices $i$ for each $\a^i$ that constitute the group $G_x$, this include the root $\a^0$ if it belongs to $G_x$. The coefficients $n_{i}^{G_x}$ are the Kac labels for $G_x$. For each $G_x$ we find out the extended Dynkin diagram by adding the root $\a^{G_x}$ with Kac label $n_{0}^{G_x}=1$. This is similar to adding $\a^0$ in the previous step. For the simple group $G_x$ that contains the root $\a^0$, we define $w_0\geq0$ as
\begin{equation}
w_0\equiv N\a^0{\cdot}a=-\sum_{i=1}^{8}n_iw_i~,
\end{equation}
where $n_i$ are the Kac labels of $E_8$, $n_i=\{2,3,4,5,6,4,2,3\}$. We also define the coefficients $w_{0}^{G_x}$ as
\begin{equation}
\sum_{i\in J_x}n_{i}^{G_x}w_i+w_{0}^{G_x}=N~.
\label{wgx}
\end{equation}
Our problem now reduces to finding the right set of integers that satisfy \Eq{wgx} and 
\begin{equation}
\sum_{I=0}^{8}n_Iw_I=0\,\,\pmod N~,
\end{equation}
where $n_I=\{n_0,n_i\}=\{1,n_i\}$.

\begin{figure}[h]
 \setlength{\unitlength}{.8mm}
\begin{center}
\begin{picture}(100,30)(58,0)
\put(20,0){\circle{5}}
\put(45,0){\circle{5}}
\put(60,0){\circle{5}}
\put(75,0){\circle{5}}
\put(90,0){\circle{5}}
\put(75,21){\circle{5}}
\put(47,0){\line(1,0){11}}
\put(62,0){\line(1,0){11}}
\put(77,0){\line(1,0){11}}
\put(75,2){\line(0,1){17}}
\put(19,-7){$\a^0$}
\put(44,-7){$\a^3$}
\put(59,-7){$\a^4$}
\put(74,-7){$\a^5$}
\put(89,-7){$\a^6$}
\put(78,24){$\a^8$}
\put(125,0){\circle{5}}
\put(150,0){\circle{5}}
\put(165,0){\circle{5}}
\put(180,0){\circle{5}}
\put(195,0){\circle{5}}
\put(180,21){\circle{5}}
\put(165,21){\circle{5}}
\put(125,21){\circle{5}}
\put(152,0){\line(1,0){11}}
\put(167,0){\line(1,0){11}}
\put(182,0){\line(1,0){11}}
\put(180,2){\line(0,1){17}}
\put(165,2){\line(0,1){17}}
\put(125,2){\line(0,1){17}}
\put(124,-7){$\a^0$}
\put(149,-7){$\a^3$}
\put(164,-7){$\a^4$}
\put(179,-7){$\a^5$}
\put(194,-7){$\a^6$}
\put(183,24){$\a^8$}
\put(118,24){$\a^{SU_2}$}
\put(157,24){$\a^{SO_{10}}$}
\put(122.1,19){\Large{$\times$}}
\put(162.2,19){\Large{$\times$}}
\put(102,10){\LARGE\boldmath$\rightarrow$}
\end{picture}
\vspace{5mm}
\caption[Extended Dynkin diagrams of $SU_2$ and $SO_{10}$]{The extended Dynkin diagram of $E_8$ with three dots removed is shown on the left. The extended Dynkin diagrams of $SU_2$ and $SO_{10}$ are shown on the right.}
\label{e8dr}
\end{center}
\end{figure}
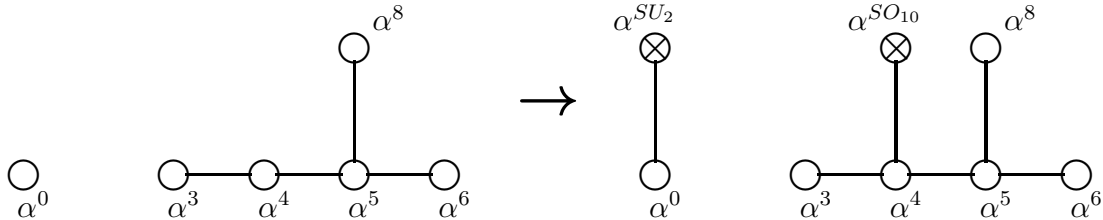

We illustrate this process by giving an example\ft{In fact, we may use the Dynkin diagram technique to verify the list of all $\ZZ_7$-invariant subgroups of $E_8$ as listed in Table~\ref{giz7s}, which we have found by the alternate, computational method described above.}. We take a group from Table~\ref{giz7s}, say, $SO_{10}\times SU_2\times U_{1}^{2}$ and look for its possible breaking with a Wilson line. One of the choices for the coefficients $s^I$ that satisfies \Eq{nisi} and breaks $E_8$ to $SO_{10}\times SU_2\times U_{1}^{2}$ is $[s_0|s_i]=[0|11000010]$. We can safely remove the roots $\a^1$, $\a^2$ and $\a^7$ from the extended Dynkin diagram of $E_8$, the resulting diagram is shown in Figure~\ref{e8dr}. We add the roots $\a^{SO_{10}}$ and $\a^{SU_2}$ and get the extended Dynkin diagram for $SO_{10}$ and $SU_2$, respectively.

The highest root of $SO_{10}$ is $(0,1,0,0,0)$ which is written as
\begin{equation}
\q^{SO_{10}}=1\a^3+2\a^4+2\a^5+1\a^8+1\a^6~,
\end{equation}
where $\a^3=(2,-1,0,0,0)$, $\a^4=(-1,2,-1,0,0)$, $\a^5=(0,-1,2,-1,-1)$, $\a^8=(0,0,-1,2,0)$ and $\a^6=(0,0,-1,0,2)$ are the simple roots of $SO_{10}$. The Kac labels are found from the metric tensors in Table -7 of \cite{slansky1}. This gives $n_{i}^{SO_{10}}=\{1,2,2,1,1\}$. For $SU_2$ group $n_{i}^{SU_{2}}=\{1\}$. We look for coefficients in the form $[\w_i|\w_0\w_{0}^{SU_2}\w_{0}^{SO_{10}}]$ that satisfy the following conditions
\begin{eqnarray}
w_0+2w_1+3w_2+4w_3+5w_4+6w_5+4w_6+2w_7+3w_8&=&0\,\pmod 7~,\nn\\
w_{0}^{SO_{10}} + w_3 + 2w_4 + 2w_5 + w_6 + w_8 &=& 7~,\nn\\
w_{0}^{SU_2} + w_0 &=& 7~.
\label{aws}
\end{eqnarray}
We use {\sl Mathematica}\TM\ and find the possible breaking pattern of $SO_{10}\times SU_2\times U_{1}^{2}$, shown in Table~\ref{t3}.
\begin{table}[ht]
\begin{center}\def\baselinestretch{1.2}\small\normalsize
\begin{tabular}{c|c}
\bs$[\w_i|\w_0\,\w_{0}^{SU_2}\,\w_{0}^{SO_{10}}]$ & \bs Group\\
\hline
$[00000000|077]$ & $SO_{10}\times SU_2$\\
$Not\,\,Possible$ & $SO_{10}$\\
$Not\,\,Possible$ & $SO_8\times SU_2$\\
$[00200000|615]$ & $SO_8$\\
$[00600100|070]$ & $SU_5\times SU_2$\\
$[00000100|346]$ & $SU_5$\\
$Not\,\,Possible$ & $SU_4\times SU_{2}^{3}$\\
$[00003001|070]$ & $SU_4\times SU_{2}^{2}$\\
$[00010000|255]$ & $SU_4\times SU_2$\\
$[00110000|524]$ & $SU_4$\\
\hline
\end{tabular}\quad\qquad
\begin{tabular}{c|c}
\bs$[\w_i|\w_0\,\w_{0}^{SU_2}\,\w_{0}^{SO_{10}}]$ & \bs Group\\
\hline
$Not\,\,Possible$ & $SU_3\times SU_{2}^{3}$\\
$[00001000|165]$ & $SU_3\times SU_{2}^{2}$\\
$[00010100|524]$ & $SU_3\times SU_2$\\
$[00110100|163]$ & $SU_3$\\
$Not \,\, Possible$ & $SU_{2}^{5}$\\
$[00201000|073]$ & $SU_{2}^{4}$\\
$[00101000|434]$ & $SU_{2}^{3}$\\
$[00111000|612]$ & $SU_{2}^{2}$\\
$[00111100|251]$ & $SU_2$\\
$Not\,\,Possible$ & $U_{1}^{8}$\\
\hline
\end{tabular}\\[2mm]
\end{center}
\caption[Table for the breaking of $SO_{10}\times  SU_2\times U_{1}^{2}$ with one Wilson line.]{Table for the breaking of $SO_{10}\times  SU_2\times U_{1}^{2}$ with one Wilson line; the $U_1$ factors are suppressed.} 
\label{t3}
\end{table}
 In the table, {\it Not Possible} means that we could not find a solution to the \Eq{aws} that would result in the corresponding group which is a subgroup of $SO_{10}\times SU_2\times U_{1}^{2}$. It is possible to get grand-unifying groups, such as $SO_{10}\times U_{1}^{3}$ or $SU_5\times U_{1}^{4}$, and Standard Model type gauge group, $SU_3\times SU_2\times U_{1}^{5}$, with an additional shift in the $E_8$ lattice. However, there are redundancies in the  final massless contents. Take for example the $E_8\rightarrow SO_{10}\times SU_2\times U_{1}^{2}\rightarrow SU_3\times SU_2\times U_{1}^{5}$ branching, with the $U_1$ charges omitted: 
\begin{equation}
 {\bf 248}\to 1({\bf 8,1})\op 1({\bf 1,3})\op 20({\bf 1,2})\op 15({\bf 3,1})\op 15({\bf \bar{3},1})\op 6({\bf 3,2})\op 6({\bf \bar{3},2})\op 35({\bf1,1})~.
\label{spec}
\end{equation}
After truncating the antichiral states, which affects only the multiplicity factors of some of the states, the massless spectrum on any one of the seven four-planes within one ten-plane (Figure~\ref{gstr}) is, somewhat formally:
\begin{equation}
 {\bf 248}~\TO~
  \inv7\big[ 1({\bf 8,1})\op 1({\bf 1,3})\op 14({\bf 1,2})\op 15({\bf 3,1})\op
   8({\bf \bar{3},1})\op 6({\bf 3,2})\op 4({\bf \bar{3},2})\op 27({\bf1,1})\big]~.
\label{spec2}
\end{equation}
That is, \Eq{spec2} gives the contribution of the indicated massless states to the various index theorems within any one of the seven four-planes. In fact, as the compact space spanned by $x_5,\cdots,x_{10}$ is unobservably small, the seven four-planes in the upper ten-plane effectively coalesce into one four-dimensional World, with the particle content indicated within the square brackets of \Eq{spec2}; the analogous holds for the seven four-planes within the lower ten-plane; see Figure~\ref{gstr}.

 Using the above procedure it is possible to find the symmetry breaking pattern of all the groups listed in Table~\ref{giz7s} in Appendix~\ref{nsr}.

\section{$\ZZ_8$ and $\ZZ_{12}$ Orbifold Models}
 \label{z8z12}
 To continue the search for four-dimensional models by compactifying $M$-theory on higher order orbifolds, one should take a look at the $S^1/\ZZ_2\times T^6/\ZZ_8$ and  $S^1/\ZZ_2\times T^6/\ZZ_{12}$ orbifolds. The action of the orbifold group elements on the compact coordinates is similar to that one in Table~\ref{gstr}, except that $\b^4$ in $\ZZ_8$, and $\b^3,\b^4$ elements in $\ZZ_{12}$, together with their $\a$-twisted versions, have additional fixed-point sets, which need to be counted. The network of intersecting fixed-point planes now includes ten,- seven,- six,- five- and four-planes, similar to those found in Refs.\cite{doran3,doran2}.
The $\ZZ_8$ and $\ZZ_{12}$ invariant groups that survive on a ten-plane can again be found using a {\sl Mathematica}\TM\ program analogous to that used for the $\ZZ_7$ orbifold. As the first step, the list of $\ZZ_8$ and $\ZZ_{12}$ shift vectors required for such computations has been constructed: There are 809 $\ZZ_8$ and 6309 $\ZZ_{12}$ vectors that satisfy the supersymmetry-preserving condition. This includes order-2, -3 and -4 shifts, which respectively generate $\ZZ_2$, $\ZZ_3$ and $\ZZ_4$ subgroups of either $\ZZ_8$ or $\ZZ_{12}$.
 
 We expect the massless spectra of four-dimensional models constructed from $S^1/\ZZ_2\times T^6/\ZZ_N$ orbifold, in the limit $x_{11}\rightarrow 0,$ to be equivalent to those obtained from $E_8\times E_8$ string theory on $T^6/\ZZ_N$ orbifolds\cite{dixon2}. In\cite{katsuki1} all the $\ZZ_N$ invariant subgroups have been listed along with the untwisted states for each of this groups. The untwisted states are obtained in a way similar to $\ZZ_7$ orbifold as explained in Section~\ref{augs}. For the twisted states we have an additional restriction, they are selected by a GSO projection operator\cite{kobayashi1}. Since the number of twisted states is too large, Ref.\cite{katsuki1} lists only the untwisted states; the $U(1)$ charges are also omitted for brevity but are computed as done above.
 
For $\ZZ_N$ orbifolds in string theory the shift vectors are chosen to preserve both the supersymmetry and modular invariance restrictions. In $M$-theory, however, a precise analogue of the latter restriction is lacking. It therefore does not follow that the orbifold models listed in Ref.\cite{katsuki1} would turn out to be the same as those models constructed by the present method. This problem is still open and needs to be explored. Considering the fact that $\ZZ_7$ orbifold models from $M$-theory and string theory turn out the same, one would expect this also for $\ZZ_8$ and $\ZZ_{12}$ orbifold models.

\section{Conclusion}
\label{con}
We have examined the synergistic application of several related techniques used to determine the massless spectrum of  $T^6/\ZZ_7\times S^1/\ZZ_2$
orbifolds in  $M$-theory. This gives us two pairs of four-plane Worlds separated by a line segment. We have specified the gauge groups and matter content of these four-dimensional Worlds.

None of the models with only untwisted states contains a four-plane World in which the massless spectrum would match the World in which we live. Typically, one finds no chiral states in anomaly free models, and/or the gauge groups do not resemble the standard model or any of the GUT group.
 
Among models where anomalies are canceled by some twisted states, the four-dimensional models are closely related to to those obtained from compactifying $E_8\times E_8$ heterotic string theory on $T^6/\ZZ_7$ orbifold~\cite{katsuki1}. In the limit $x_{11}\to0$, the $M$-theory orbifold models should have the same massless spectra as the corresponding string theory orbifolds. We do not, at this stage, know the details of this limit, foremost because $M$-theory---unlike string theory---does not have a well-understood ``fundamental'' description. However, our considerations do provide the necessary twisted states to cancel the anomalies, which also introduces some chiral states in the massless spectrum. In our models, five-dimensional planes connect pairs of four-plane Worlds, which are otherwise separated from each other. Since there are no anomalies on these five-planes, possible additional gauge symmetries in these planes cannot be determined in this way\cite{doran3,doran2,kaplunovsky1,doran1}. In some of our models, some of the twisted states extend through these five-planes; the other twisted states need not extend in this way and may be localized to the respective four-plane Worlds. This is an open problem and needs further insight.

Our calculations in finding the $\ZZ_7$ invariant subgroups of $E_8$ heavily relies on {\sl Mathematica}\TM\ programs, and the procedure can be applied for any other orbifold group in general. The programming aspects of the relevant {\sl Mathematica}\TM\ code will be discussed in a separate report. One might think that orbifolding with higher order groups, like $\ZZ_8$ or $\ZZ_{12}$ would give us more realistic and smaller groups. This much is indeed true, but the number of twisted states in that case grows too fast for realistic application~\cite{katsuki1}; these orbifolds do not appear to be viable as potential candidates to relate $M$-theory with the real World. However, this is just an assumption, a complete calculation is necessary to arrive at such statement. Initially we started our research with $\ZZ_N$ orbifold in mind and we have partially worked on $\ZZ_8$ and $\ZZ_{12}$. Getting the untwisted states for any of these orbifolds was easy, but for the twisted states we had to rely on string theory. Even though we have a general procedure (in principle) to find out the twisted states, practically the calculation needs some computational tools. As presently implemented, our {\sl Mathematica}\TM\ code for finding the twisted states has a significant limitation: it works for any string theory models with up to two $U(1)$ factors. The obvious need for generalizations in this direction provides goals for further research.

We have used Wilson lines to break the four-dimensional gauge group to a more realistic standard-like model. The example we have presented gives the gauge groups on one of the upper or lower four-plane Worlds. For a complete list of models, we need to find the symmetry breaking on the opposite four-plane World, take the $x_{11}\to0$ limit,  calculate the anomalies, and ensure their cancellation. Although this leads to a plethora  of models, the symmetry groups on the four-planes are much smaller than the ones obtained without employing Wilson lines.

Finally, there are mechanisms in string theory to break the gauge symmetry in a way that reduces the rank of the gauge group~\cite{casas1,casas2,font1}. The precise analogue of this mechanism is lacking in $M$-theory context. From the above indications however, one would expect an $M$-theory mechanism to exist, such that its $x_{11}\to0$ limit (with the more detailed dynamics taken into account), is that of Refs~\cite{casas1,casas2,font1}. This remains an open question.

\vfill
\bigskip\paragraph{\bfseries Acknowledgments:}
We should like to thank Mike Faux and Tatsuo Kobayashi for helpful discussions, and The U.S.\ Department of Energy for their support under the grant DE-FG02-94ER-40854.\vfill

\clearpage
\appendix
\clearpage

\section{Subgroups of $E_8$}
\label{sge8}
We find out all the subgroups of $E_8$ using Dynkin Diagram\cite{wybourne1}. The procedure is to remove circles in all possible ways from the  extended Dynkin diagram and read off the subgroup from the remaining circles. In this way, we get a total of 99 subgroups of $E_8$ given in Table~\ref{atsg}.
\renewcommand{\arraystretch}{1.3}
\setlength{\LTcapwidth}{6.5in}
\begin{longtable}{|c|c|c|c|c|c|c|c|c|c|c|c|}
\multicolumn{1}{c}{  } &
\multicolumn{1}{c}{\bf Subgroups} &
\multicolumn{1}{c}{$\bm k$} &
\multicolumn{1}{c}{$\bm m$} &
\multicolumn{1}{c}{$\bm r$} &
\multicolumn{1}{c}{$\bm o$} &
\multicolumn{1}{c}{  } &
\multicolumn{1}{c}{\bf Subgroups} &
\multicolumn{1}{c}{$\bm k$} &
\multicolumn{1}{c}{$\bm m$} &
\multicolumn{1}{c}{$\bm r$} &
\multicolumn{1}{c}{$\bm o$} \\
\hline
\endfirsthead
\multicolumn{1}{c}{  } &
\multicolumn{1}{c}{\bf Subgroups} &
\multicolumn{1}{c}{$\bm k$} &
\multicolumn{1}{c}{$\bm m$} &
\multicolumn{1}{c}{$\bm r$} &
\multicolumn{1}{c}{$\bm o$} &
\multicolumn{1}{c}{  } &
\multicolumn{1}{c}{\bf Subgroups} &
\multicolumn{1}{c}{$\bm k$} &
\multicolumn{1}{c}{$\bm m$} &
\multicolumn{1}{c}{$\bm r$} &
\multicolumn{1}{c}{$\bm o$} \\
\hline
\endhead
\multicolumn{12}{r}{\small {\it Continued}}\\
\endfoot
\multicolumn{12}{r}{\hspace{2cm}}
\endlastfoot
1&$E_8$&120&0&8&&51&$SO_8\times SU_2^2$&14&2&6&$\Chekk$\\\hline
2&$E_7\times SU_2$&64&1&8&&52&$SU_5\times SU_3\times SU_2$&14&1&7&$\Chekk$\\\hline
3&$E_7$&63&0&7&$\Chekk$&53&$SU_5\times SU_2^4$&14&4&8&$\Chekk$\\\hline
4&$SO_{16}$&56&0&8&&54&$SU_4^2\times SU_2^2$&14&2&8&$\Chekk$\\\hline
5&$SO_{14}\times SU_2$&43&1&8&&55&$SO_8\times SU_2$&13&1&5&\\\hline
6&$SO_{14}$&42&0&7&$\Chekk$&56&$SU_5\times SU_3$&13&0&6&\\\hline
7&$E_6\times SU_3$&39&0&8&&57&$SU_5\times SU_2^3$&13&3&7&\\\hline
8&$E_6\times SU_2^2$&38&2&8&&58&$SU_4^2\times SU_2$&13&1&7&\\\hline
9&$E_6\times SU_2$&37&1&7&$\Chekk$&59&$SU_4\times SU_3^2\times SU_2$&13&1&8&\\\hline
10&$E_6$&36&0&6&$\Chekk$&60&$SO_8$&12&0&4&\\\hline
11&$SU_9$&36&0&8&$\Chekk$&61&$SU_5\times SU_2^2$&12&2&6&\\\hline
12&$SO_{12}\times SU_3$&33&0&8&&62&$SU_4^2$&12&0&6&\\\hline
13&$SO_{12}\times SU_2^2$&32&2&8&&63&$SU_4\times SU_3^2$&12&0&7&\\\hline
14&$SO_{12}\times SU_2$&31&1&7&&64&$SU_4\times SU_3\times SU_2^3$&12&3&8&\\\hline
15&$SO_{12}$&30&0&6&$\Chekk$&65&$SU_3^4$&12&0&8&\\\hline
16&$SU_8\times SU_2$&29&1&8&&66&$SU_5\times SU_2$&11&1&5&\\\hline
17&$SU_8$&28&0&7&$\Chekk$&67&$SU_4\times SU_3\times SU_2^2$&11&2&7&\\\hline
18&$SO_{10}\times SU_4$&26&0&8&&68&$SU_4\times SU_2^5$&11&5&8&\\\hline
19&$SO_{10}\times SU_3\times SU_2$&24&1&8&&69&$SU_3^3\times SU_2^2$&11&2&8&\\\hline
20&$SU_7\times SU_3$&24&0&8&&70&$SU_5$&10&0&4&\\\hline
21&$SO_{10}\times SU_3$&23&0&7&$\Chekk$&71&$SU_4\times SU_3\times SU_2$&10&1&6&\\\hline
22&$SO_{10}\times SU_2^3$&23&3&8&$\Chekk$&72&$SU_4\times SU_2^4$&10&4&7&\\\hline
23&$SU_7\times SU_2^2$&23&2&8&$\Chekk$&73&$SU_3^3\times SU_2$&10&1&7&\\\hline
24&$SO_{10}\times SU_2^2$&22&2&7&$\Chekk$&74&$SU_3^2\times SU_2^4$&10&4&8&\\\hline
25&$SO_8\times SU_5$&22&0&8&$\Chekk$&75&$SU_4\times SU_3$&9&0&5&\\\hline
26&$SU_7\times SU_2$&22&1&7&$\Chekk$&76&$SU_4\times SU_2^3$&9&3&6&\\\hline
27&$SO_{10}\times SU_2$&21&1&6&$\Chekk$&77&$SU_3^3$&9&0&6&\\\hline
28&$SU_7$&21&0&6&$\Chekk$&78&$SU_3^2\times SU_2^3$&9&3&7&\\\hline
29&$SU_6\times SU_4$&21&0&8&$\Chekk$&79&$SU_3\times SU_2^6$&9&6&8&\\\hline
30&$SO_{10}$&20&0&5&&80&$SU_4\times SU_2^2$&8&2&5&\\\hline
31&$SU_5^2$&20&0&8&&81&$SU_3^2\times SU_2^2$&8&2&6&\\\hline
32&$SO_8\times SU_4\times SU_2$&19&1&8&&82&$SU_3\times SU_2^5$&8&5&7&\\\hline
33&$SU_6\times SU_3\times SU_2$&19&1&8&&83&$SU_2^8$&8&8&8&\\\hline
34&$SO_8\times SU_4$&18&0&7&&84&$SU_4\times SU_2$&7&1&4&\\\hline
35&$SO_8\times SU_3^2$&18&0&8&&85&$SU_3^2\times SU_2$&7&1&5&\\\hline
36&$SU_6\times SU_3$&18&0&7&&86&$SU_3\times SU_2^4$&7&4&6&\\\hline
37&$SU_6\times SU_2^3$&18&3&8&&87&$SU_2^7$&7&7&7&\\\hline
38&$SO_8\times SU_3\times SU_2^2$&17&2&8&&88&$SU_4$&6&0&3&\\\hline
39&$SU_6\times SU_2^2$&17&2&7&&89&$SU_3^2$&6&0&4&\\\hline
40&$SU_5\times SU_4\times SU_2$&17&1&8&&90&$SU_3\times SU_2^3$&6&3&5&\\\hline
41&$SO_8\times SU_3\times SU_2$&16&1&7&$\Chekk$&91&$SU_2^6$&6&6&6&\\\hline
42&$SO_8\times SU_2^4$&16&4&8&$\Chekk$&92&$SU_3\times SU_2^2$&5&2&4&\\\hline
43&$SU_6\times SU_2$&16&1&6&$\Chekk$&93&$SU_2^5$&5&5&5&\\\hline
44&$SU_5\times SU_4$&16&0&7&$\Chekk$&94&$SU_3\times SU_2$&4&1&3&\\\hline
45&$SU_5\times SU_3^2$&16&0&8&$\Chekk$&95&$SU_2^4$&4&4&4&\\\hline
46&$SO_8\times SU_3$&15&0&6&$\Chekk$&96&$SU_3$&3&0&2&\\\hline
47&$SO_8\times SU_2^3$&15&3&7&$\Chekk$&97&$SU_2^3$&3&3&3&\\\hline
48&$SU_6$&15&0&5&$\Chekk$&98&$SU_2^2$&2&2&2&\\\hline
49&$SU_5\times SU_3\times SU_2^2$&15&2&8&$\Chekk$&99&$SU_2$&1&1&1&\\\hline
50&$SU_4^2\times SU_3$&15&0&8&$\Chekk$&100&$U_1^8$&0&0&8&\\\hline
\caption{All the subgroups of $E_8$ and the values of the identifiers. The entries under the headings $k$, $m$ and $r$ are defined in Eqs.\eq{ekV} and \eq{M1}, respectively. Only the subgroups labeled by a check-mark in the ``$o$'' column may occur in our $\ZZ_7$-orbifold models.}
\label{atsg}
\end{longtable}
\vspace{-15mm}

\section{$\ZZ_7$-Invariant Subgroups of $E_8$ and Generalizations}
\label{nsr}
Among the subgroups of $E_8$, we are only interested in those that are marked with a check in column `$o$' of Table~\ref{atsg}. These are the possible candidates as $\ZZ_7$ invariant subgroups. We arrive at this conclusion by finding all the values of $k_{\bm V}$\eq{kv}, each of which is the number of $E_8$ roots that satisfy the condition $P^I{\cdot}V^I\in\ZZ$ for a given shift vector $V^I$, with $V^I$ ranging over all $\ZZ_7$ shift vectors. This narrows down our choices to 30 subgroups of $E_8$.

Next, we calculate the values of $m$ (number of  $SU_2$ factors in a group) for all of these choices, this we do manually from Table~\ref{atsg}. If $k$ and $m$ do not specify the group we find the values of $r$ (rank of a group), also manually, from the table.

Finally, we use {\sl Mathematica}\TM\ to calculate the values of the identifiers $k$, $m$ and $r$ from the surviving roots of $E_8$ and compare them with the  values in Table~\ref{atsg} to determine the $\ZZ_7$ invariant subgroups of $E_8$. This reduces our choices to 14 subgroups only.
\begin{table}[ht]
\begin{center}
\begin{tabular}{|c|c|}
  \multicolumn{1}{c}{} & \multicolumn{1}{c}{\bs Subgroup} \\\hline
 1 & $E_7$ \\\hline
 2 & $E_6\times SU_2$ \\\hline
 3 & $E_6$ \\\hline
 4 & $SO_{14}$ \\\hline
\end{tabular}
 \quad
\begin{tabular}{|c|c|}
  \multicolumn{1}{c}{} & \multicolumn{1}{c}{\bs Subgroup} \\\hline
 5 & $SO_{12}$ \\\hline
 6 & $SO_{10}\times SU_3$ \\\hline
 7 & $SO_{10}\times SU_2$ \\\hline
 8 & $SO_8\times SU_3$ \\\hline
\end{tabular}
 \quad
\begin{tabular}{|c|c|}
  \multicolumn{1}{c}{} & \multicolumn{1}{c}{\bs Subgroup} \\\hline
 9  & $SU_8$ \\\hline
 10 & $SU_7\times SU_2$ \\\hline
 11 & $SU_7$ \\\hline
 12 & $SU_6\times SU_2$\\\hline
\end{tabular}
 \quad
\begin{tabular}{|c|c|}
  \multicolumn{1}{c}{} & \multicolumn{1}{c}{\bs Subgroup} \\\hline
 13 & $SU_5\times SU_4$ \\\hline
 14 & $SU_5\times SU_3\times SU_2$ \\\hline
 \multicolumn{2}{c}{}\\
 \multicolumn{2}{c}{}\\
\end{tabular}
\\[2mm]
\end{center}
\caption{Subgroups of $E_8$ invariant under $\ZZ_7$ shifts.} 
\label{giz7s}
\end{table}

\noindent
{\bs A closing remark:} When considering higher-order orbifolds, the above identifiers, $k,m,r$ will not identify the subgroups uniquely. However, the counting of the $SU(2)$ factors can be generalized in a very simple way as follows. The $SU(n)$ groups are uniquely characterized by the fact that they have $n\choose2$ positive roots that form a Pascal triangle:
\begin{equation}
\begin{gathered}
 E_{\a^1+\a^2+\cdots+\a^{n-1}}\quad\\
 \cdots\qquad\cdots\qquad\cdots\quad\\
 \cdots\qquad\cdots\qquad\cdots\qquad\cdots\quad\\
 E_{\a^1+\a^2} ~~ E_{\a^2+\a^3} ~~\cdots~~ E_{\a^{n-2}+\a^{n-1}}\quad \\
 E_{\a^1} \qquad E_{\a^2} \qquad E_{\a^3} \quad\cdots\quad E_{\a^{n-2}} \qquad E_{\a^{n-1}}
\end{gathered}
\end{equation}
where $E_{\a^i+\a^j}=[E_{\a^i},E_{\a^j}]$, \etc\ 
This implies an analogous system of relations for the vectors $z^i$, to be found among the relations such as Eqs.\eq{rrln}. One then first identifies all the $SU(2)$ sub-systems in this way, and then all other $SU(n)$ factors, increasing $n$ incrementally, one by by one, from $n=3$ and until the subgroup of $E_8$ has been identified.
 The corresponding number of $SU(n)$ factors may then be labeled $m_n$, where $m_2$ is the $m$ as defined in Eqs.\eq{M1} and\eq{M2}.

The special treatment of the $SU(n)$ subgroups was simply for illustration purposes; other Lie groups also have characteristic positive root systems, encoding the correspondingly characteristic commutation relations among the positive roots. The $SU(2)$ subgroup factors remain, of course, the simplest to find.

 In general then, for any system of $\G$-invariant roots $z^i$, we find the complete system of ternary relations such as Eqs.\eq{rrln}. Within this system, we identify a sub-system of relations as corresponding to the characteristic commutation relations of a Lie group $H$. This identifies $H$ as factor in the maximal $\G$-invariant subgroup of $E_8$.

\clearpage
\bibliographystyle{jhepm}
\bibliography{ahsan_ref}
\end{document}